\documentclass[aps, prd, amsmath, floats, floatfix, onecolumn, nofootinbib]{revtex4-2}
\pdfoutput=1
\usepackage{graphicx}
\usepackage[dvipsnames]{xcolor}
\usepackage{latexsym}

\usepackage{amssymb}
\usepackage{amsmath}
\usepackage{enumitem}
\usepackage[bookmarksnumbered=true]{hyperref} 
\hypersetup{
    colorlinks = true,
    linkcolor = blue,
    anchorcolor = blue,
    citecolor = blue,
    filecolor = blue,
    urlcolor = blue
    }

\newcommand{\be}{\begin{eqnarray}}
\newcommand{\ee}{\end{eqnarray}}

\newcommand{\bea}{\begin{eqnarray}}
\newcommand{\eea}{\end{eqnarray}}

\begin{document}

\title{Do Black Holes Exist?}

\author{Leonardo Modesto}
\email{leonardo.modesto@unica.it}

\author{Edoardo Rattu}
\email{edoardo.rattu03@gmail.com}

\affiliation{Dipartimento di Fisica, Universit\`a di Cagliari, Cittadella Universitaria, 09042 Monserrato, Italy}
\affiliation{I.N.F.N, Sezione di Cagliari, Cittadella Universitaria, 09042 Monserrato, Italy}


\begin{abstract}
We carefully investigate, extend, and shed new light on the McVittie exact solution of Einstein's gravity (EG) with the focus on the implications in the Universe we live in. It turns out that the only known exact solution of EG, which interpolates between an asymptotic homogeneous and isotropic Universe 
and a {\em Schwarzschild black hole}, is actually singular in $2M$, namely the curvature invariants diverge and the spacetime is geodetically incomplete in $2M$. 
Very important: all energy conditions are satisfied beside the dominant one (DEC) that is violated inside the radius $8M/3$, which is of the order but bigger than $2M$. 
Notice that $2M$ is not the event horizon, but a curvature singularity 
(not a coordinate singularity like for the Schwarzschild spacetime).
covered by an apparent horizon that at the actual stage of the Universe nearly coincides with $2M$. 
Moreover, the curvature singularity is not analytic with respect to the dynamics of the Universe encoded in the Hubble function $H(t)$: for arbitrarily small but not zero $H^\prime(t)$ the curvature invariants are singular, while for $H^\prime(t)$ identically zero they are regular. Therefore, we can not analytically decouple the black hole from the entire Cosmos, namely 
we can not assume the Schwarzschild solution locally and the FRW metric at large scale without violating the analyticity of the metric.

Since the spacetime does not exist for $r \leqslant 2M$ because the curvature invariants are defined only 
for $r >2M$, and since the DEC is violated for $r<8M/3$, we are allowed to doubt the existence of black holes in our Universe as understood up to now. In particular, we expect the violation of DEC to be catastrophic for the spacetime stability below $8M/3$. 

We also build and study a toy model for the gravitational collapse generalizing the Vaidya to the McVittie-Vaidya metric. 
Although dynamical, the singularities remain in the same locations. 
In order to achieve the curvature smoothness and geodesic completion, we propose a simple solution of both the singularities in $2M$ and in $t=0$ in Conformal Einstein's gravity. The spacetime turns out to be geodetically complete and the singularities become asymptotic unreachable regions of the spacetime.

Finally, we address the singularity issue in $2M$ allowing the violation of all energy conditions. The outcome is a giant soliton, but no black hole.

\end{abstract}

\maketitle

\section*{Introduction}
We investigate whether Einstein's gravity predicts the presence of black holes in our Universe. 
Indeed, it is common to assume the spacetime to be asymptotically flat when deriving the metric of a compact object consistent with spherical or axial symmetry. However, the Universe we live in is not asymptotically Minkowski but is instead described by the Freedman-Robertson-Walker (FRW) metric.
We will show that the above mathematical simplification is not correct  
 whether we want to describe the local physics in a dynamical Universe.
It is commonly said that we can consider the spacetime asymptotically flat because the scale of the Universe is vastly larger than that of a compact object or a black hole. We will see in this paper that this statement is incorrect. Indeed, the metric of a black hole, actually a black hole mimicker because singular in $2M$, embedded in an expanding Universe is not analytic in the derivative of the Hubble function $H(t)$. 
In other words, the transition from nearly constant $H(t)$ to constant determines a transition between a metric singular in $2M$ and the Schwarzschild (or Schwarzschild - (anti)de Sitter) spacetime, which is smooth for every $r>0$.

In order to investigate this issue, we will review and expand about the McVittie exact solution of Einstein's gravity \cite{OriginalMV}
consistently with the positivity of the energy conditions
%
\footnote{As we will see later, this is not exactly true, and this will cast some doubts upon the plausibility of the solution near and above $2M$.}. 

The paper is organized as follows. In the first section we will review the McVittie metric in two different coordinate systems providing the explicit form of the energy momentum tensor. We will evaluate several curvature invariants 
such as the Ricci scalar, the Ricci tensor square, the Kretschmann tensor, and the Weyl tensor.
We will solve the geodesic equation near $2M$ showing that the spacetime is geodetically incomplete. In section two we propose a Vaydia -  McVittie metric as a toy model for the gravitational collapse in an expanding Universe and we will show that all the singularities stay in the same location. 
In section three we will perform a conformal rescaling of McVittie metric, and we will show that the singularities in $r = 2M$ and in $t = 0$ can be moved out of our Universe. In section four, we will take a different approach and propose a profile $M(r)$ to avoid the singularity in $2M$. We will study the energy conditions and the presence of apparent horizons before and after the replacement $M(r)$ into the metric. Finally, we will dedicate to conclusions and remarks.

We will adopt the metric signature $\{-, +, +, +\}$ and the units such that, in particular, $c = 1$ and $G = 1$ (Planck's units). 

\section{Review of the McVittie metric}

In 1933, G. C. McVittie proposed an exact solution of EG describing a massive object embedded in an expanding Universe \cite{OriginalMV}. The author assumed local spherical symmetry around the massive body, 
 the absence of any physical process involving fluxes of matter, 
 and eventually an isotropic pressure in every point of the Universe. 
 Therefore, starting from Einstein's Field Equations, which can be written in the form
\begin{equation}
    R_{\mu\nu} = {8\pi }  \left(T_{\mu\nu}-\frac{1}{2}g_{\mu\nu}T\right),
    \label{equazioni_del_moto_di_Einstein_usate_per_il_tensore_energia_impulso}
\end{equation}
McVittie was able to find the following exact solution for the metric,
\begin{equation}
\begin{split}
    ds^2 = -\left({{1-\frac{M}{2a(t)R}}\over{1+\frac{M}{2a(t)R}}}\right)^2dt^2+\left( 1+\frac{M}{2a(t)R}\right)^4a^2(t)\left\{dR^2+R^2d\Omega^2\right\} \, , 
    \label{metrica_di_McVittie}
    \end{split}
\end{equation}
where 
$a(t)$ is the scale factor of the expanding Universe. 
If we consider the areal radius 
\begin{equation}
    r = a(t)R\left\{1+{M\over {2a(t)R}}\right\}^2 \, ,
    \label{raggio_areale} 
\end{equation}
we can recast the metric in the following form (also considered, for instance, in \cite{MatchingMV}): 
\begin{equation}
\begin{split}
    ds^2 = -\left(1-{\frac{2M}{r}}-H^2r^2\right) dt^2-{\frac{2Hr}{\sqrt{1-\frac{2M}{r}}}}drdt+{\frac{dr^2}{1-\frac{2M}{r}}}+r^2d\Omega^2 \, , 
    \label{metrica_di_McVittie_con_il_raggio_areale}
    \end{split}
\end{equation}
where $H = H(t)$ is the Hubble parameter, namely  
$H(t) = a'(t)/a(t)=\alpha/t$ for $a(t) = t^\alpha$ (prime $'$ denotes the derivative with respect to $t$). 

It should be noticed that the metric is not defined for $r \le 2M$ because of the square root appearing in the denominator of the second term. Therefore, the spacetime is defined only for $r > 2M$, i.e. there is no interior contrary to the Schwarzschild solution. 
Later we will study the causality properties of the metric and we will derive the radial coordinate distance of the apparent horizon from $2M$. 

From the Field Equations by reversing engineering, 
it is possible to determine the components of the energy-momentum tensor $T_\mu^\mu$. 
In particular, 
the density $\rho$ and the pressure $p$ are:
\be
&&
    \rho = -T^t_t =  -\left( g^{tt}T_{tt}+g^{tr}T_{rt}\right) =  \frac{3}{8\pi}H^2
    \label{formuladensita}  \, , \\
&&
    p = T^r_r = g^{rr}T_{rr} + g^{rt}T_{tr} = -\frac{1}{8\pi}\left( 3H^2+\frac{2H_t}{\sqrt{1-\frac{2M}{r}}}\right) ,
    \label{equazione_per_la_pressione_con_H_implicito} 
\ee
where by $H_t$ we defined the derivative of $H$ with respect to $t$.
 The density (\ref{formuladensita}) is clearly always positive or zero. 
 Assuming a matter dominated Universe, i.e. 
 $a(t)\propto$ $t^{\frac{2}{3}}$, so that $H(t) \propto \frac{2}{3t}$, the expression for $p$ becomes:
\begin{equation}
    p = \rho\left( \frac{1}{\sqrt{1-\frac{2M}{r}}}-1\right) .
    \label{formula_pressione}
\end{equation}
From this latter expression, we see that $p$ is positive for $r>2M$, while it  
diverges near $r = 2M$ (see Fig.\ref{figurespressureandDEC} for three different values of $t$). 

\subsection{Energy Conditions}
\begin{figure}[h]
    \centering
    \includegraphics[width=0.45\linewidth]{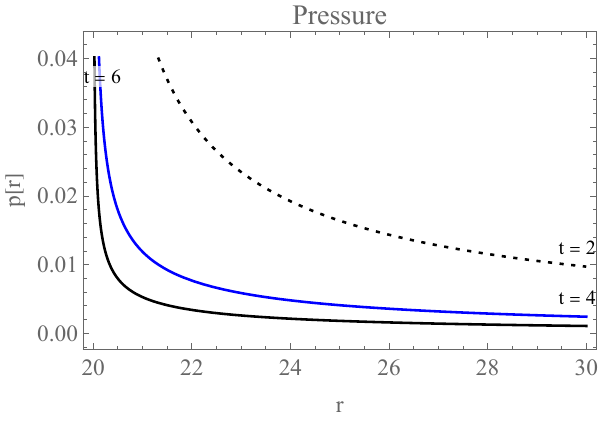}
    \hspace{1cm}
    \includegraphics[width=0.46\linewidth]{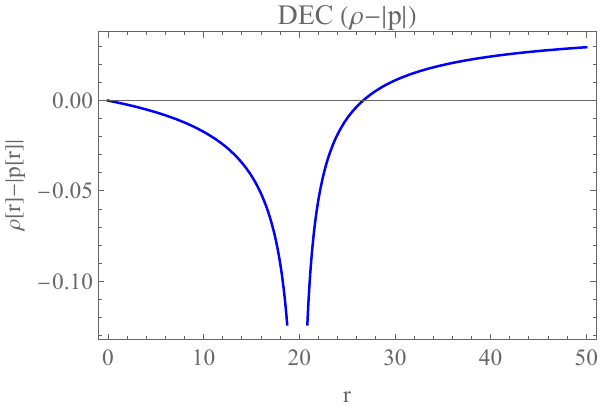}
    \caption{The plots on the left represents the pressure for three different values of $t$. The presssure $p$ is positive for $r>2M$ and diverges as $r$ tends to $2M$. The plot on the right shows the DEC and its violation for all values of $r$ lower than a certain radius, namely $8M/3$ as explained in the main text. In the the plots we made the choice $M = 10$.}
    \label{figurespressureandDEC}
\end{figure}

In this subsection, we study the following {\em energy conditions}: the WEC, the NEC, the SEC and the DEC. 
It turns out that the WEC, the NEC, and the SEC are all satisfied for $r>2M$, but the DEC is not. More precisely, the DEC is satisfied only from a certain value of the radius 
(see the plot on the right in Fig.\ref{figurespressureandDEC}), which analytically reads: 
\be
r_{\rm v} = \frac{8}{3}M > 2 M\, , 
\label{errev}
\ee
The latter value (\ref{errev}) is obtained by imposing $\rho - |p| > 0$, making use of \eqref{formula_pressione} and imposing that the quantity inside the brackets is equal to the unity. 
We would like to remark that the value $\frac{8}{3}M$ depends on the hypothesis made on $H(t)$, namely \eqref{formula_pressione}, which is consistent with a Universe dominated by dust. 
%
%
%
\subsection{Curvature Singularities}
The metric \eqref{metrica_di_McVittie_con_il_raggio_areale} shows at least two singularities, namely in $r = 0$ and $r = 2M$ 
However, $H(t)$ hides another singularity in $t = 0$ because $H$ $\propto 1/t$. Thus there are in total three potential singularities. In order to understand whether they are curvature singularities or coordinate singularities, we evaluate several curvature invariants. The simplest curvature invariant is the Ricci scalar, 
\begin{equation}
    R=\frac{6 a''(t)}{a(t) \sqrt{1-\frac{2 M}{r}}}-\frac{6 a'(t)^2}{a(t)^2 \sqrt{1-\frac{2 M}{r}}}+\frac{12
   a'(t)^2}{a(t)^2}.
\end{equation}
The Kretschmann scalar reads:
\be
   R_{\mu\nu\gamma\delta}R^{\mu\nu\gamma\delta} & = & -\frac{12 r a''(t)^2}{a(t)^2 (2 M-r)}+\frac{24 r \sqrt{1-\frac{2 M}{r}} a'(t)^4}{a(t)^4 (2 M-r)} 
   - \frac{36 r
   a'(t)^4}{a(t)^4 (2 M-r)}+\frac{48 M a'(t)^4}{a(t)^4 (2 M-r)} \nonumber \\
   &&
   -\frac{24 r \sqrt{1-\frac{2 M}{r}} a'(t)^2
   a''(t)}{a(t)^3 (2 M-r)}+\frac{24 r a'(t)^2 a''(t)}{a(t)^3 (2 M-r)}+\frac{48 M^2}{r^6}.
   \label{Kre}
   \ee
The Ricci tensor square is:
\be
   R_{\mu\nu}R^{\mu\nu} & = & -\frac{12 r a''(t)^2}{a(t)^2 (2 M-r)}+\frac{36 r \sqrt{1-\frac{2 M}{r}} a'(t)^4}{a(t)^4 (2 M-r)}
   -\frac{48 r
   a'(t)^4}{a(t)^4 (2 M-r)}+\frac{72 M a'(t)^4}{a(t)^4 (2 M-r)} \nonumber \\
   &&
   -\frac{36 r \sqrt{1-\frac{2 M}{r}} a'(t)^2
   a''(t)}{a(t)^3 (2 M-r)}+\frac{24 r a'(t)^2 a''(t)}{a(t)^3 (2 M-r)}.
   \ee
Finally, the Weyl square curvature invariant reads:
\begin{equation}
W^2 = \frac{48M^2}{r^6}.
    \label{weyl_quadro_senza_rescaling}
\end{equation}
In order the to check the correctness of the expressions above, we can simply consider $a(t) = {\rm constant}$ and observe that $R$ and $R_{\mu\nu}R^{\mu\nu}$ vanish, while the Kretschmann scalar assumes the same value of $W^2$. Indeed, the McVittie metric reduces to the Schwarzschild metric for $a^\prime(t)=0$. 

The expressions displayed above (Weyl tensor aside) lack of simplicity and clarity. Hence, it is useful to put beside them the equivalent expressions in terms of $H(t)$. 
In terms of the latter function, the curvature scalar reads:
\begin{equation}
    R=\frac{6 H'(t)}{\sqrt{1-\frac{2 M}{r}}}+12 H(t)^2\, ,
\end{equation}
the Kretschmann curvature invariant is:
\be
   R_{\mu\nu\gamma\delta}R^{\mu\nu\gamma\delta} = 
   \frac{24 H(t)^2 H'(t)}{\sqrt{1-\frac{2 M}{r}}}+\frac{12 r H'(t)^2}{r-2 M}+24 H(t)^4+\frac{48 M^2}{r^6}\, , 
   \label{Kre2}
   \ee
and the Ricci tensor square simplifies to:
\be
   R_{\mu\nu}R^{\mu\nu}  = \frac{36 H(t)^2 H'(t)}{\sqrt{1-\frac{2 M}{r}}}+\frac{12 r H'(t)^2}{r-2 M}+36 H(t)^4\, .
   \ee
If $H$ is a constant, say $H_0$, the metric \eqref{metrica_di_McVittie_con_il_raggio_areale} reduces to the Schwarzschild-de Sitter solution, whose scale factor reads $a(t) \propto \exp {H_0 t}$. 
Looking at the curvature invariants, the singularity in $2 M$ disappears for the de Sitter metric.

%
Coming back to the general McVittie metric, all the curvature invariants, with the exception of the Weyl square, are singular in $r = 2M$. Therefore, the singularity at the origin is now not relevant at all because in order to reach $r=0$ one should first cross $2M$, which is already singular. Moreover, the metric is only defined for $r > 2M$, as evident from the square root present in the curvature invariants. Hence, all the papers about the cosmological coupling of regular black holes have to be revisited according to this crucial feature \cite{cadoni, Cadoni:2024jxy, Cadoni:2023lqe,Cadoni:2023lum}.

A deeper insight shows that the singularity in $2M$ always appears coupled with 
%
the derivative of the Hubble function $H(t)$. In other words, the singularity disappears only for 
$H(t) = {\rm const.}$, namely when the Universe is asymptotically Minkowski or (Anti)de Sitter. This is tantamount to say that the spacetime is not analytic in $H^\prime(t)$. Indeed, if we define a general curvature invariant by $\mathcal{I}$, we get a discontinuity, namely:
\be
&& \mathcal{I} \rightarrow \infty \quad \forall \quad H^\prime \neq 0  \quad (H \neq {\rm const.}) \, ,  \nonumber \\
&& \mathcal{I}  < \infty \quad \mbox{for} \quad H^\prime = 0  \quad (H = {\rm const.})
\, .
\ee
This property of the solution excludes the existence of black holes as commonly understood, and they have to be replaced by objects singular (quasi naked, as we will see later) in $2M$. Furthermore, since the dominant energy condition is violated near $2M$ (\ref{errev}), we are allowed to question the stability of the solution near the singularity. 
Once again, we can not make the perturbative assumption that $H^\prime$ is very small because for any not identically vanishing value of $H^\prime$ the metric is singular in $2M$. In other words, under very plausible and conservative assumptions in EG, we can not have a Schwarzschild black hole in our Universe, but only a black hole mimicker singular in $2M$. This is a genuine prediction of EG.

One could think to take seriously the exact Schwarzschild solution in a void region of the Universe and the FRW solution outside it. However, the whole metric could not be analytic everywhere but a kind of shell should be introduced at the boundary of the void (see \cite{Stephani}). Therefore, so far the only analytic exact solution interpolating between a black hole mimicker and an asymptotically homogeneous Universe is singular in $2M$.

%

Finally, let us consider the cosmological singularity. 
If we explicitly replace $H =\alpha/t$, the Kretschmann invariant reads:
\begin{equation}
   R_{\mu\nu\gamma\delta}R^{\mu\nu\gamma\delta}= 12\left(\frac{2\alpha^4}{t^4}-\frac{2\alpha^3}{t^4\sqrt{1-\frac{2M}{r}}}+\frac{\alpha^2r}{t^4\left(r-2M\right)}+\frac{4M^2}{r^6}\right),
    \label{KretschmannHesplicito}
\end{equation}
which is singular in $t = 0$. 
Notice that if $M\neq 0$ the cosmological singularity and the singularity in $2M$ appear together in the second and the third term of (\ref{KretschmannHesplicito}), while the singularities in the first and the fourth term are respectively purely cosmological and Schwarzschild-like.

\subsection{Geodesic incompleteness}\label{geoin}
Now we investigate whether the solution \eqref{metrica_di_McVittie_con_il_raggio_areale} is geodetically complete in $r = 2M$. In particular, we focus on a photon moving radially towards the singularity. The geodesic equation deprived of the angular term ($\theta$ and $\phi$ are both constant) reads:
\begin{equation}
    {\frac{d^2r}{d\lambda^2}} + \Gamma^r_{kl}{\frac{dx^k}{d\lambda}}{\frac{dx^l}{d\lambda}} = 0,
    \label{equazione_delle_geodetiche_radiale_generale}
\end{equation}
where $\lambda$ is the affine parameter 
for the massless particle, whilst the indexes $k$ and $l$ assume the role of $r$ or $t$. 
We choose the following notation for the derivatives with respect to $\lambda$, 
\be
\frac{dr}{d\lambda}\equiv \dot r \quad {\rm and} \quad \frac{dt}{d\lambda}\equiv \dot t. 
\ee
For photons $ds^2 = 0$, thus from (\ref{metrica_di_McVittie_con_il_raggio_areale}) we get:
\begin{equation}
    {\frac{dt}{dr}} = {\frac{\dot t}{\dot r}} = -{\frac{1}{\sqrt{1-{\frac{2M}{r}}}\left(\sqrt{1-{\frac{2M}{r}}}-Hr\right)}} \equiv f,
    \label{equazione_pendenza_cono_luce}. 
\end{equation}
Combining expression (\ref{equazione_pendenza_cono_luce}) with \eqref{equazione_delle_geodetiche_radiale_generale}, we can get the following equation for the radial motion,
\begin{equation}
    \ddot r + \left( \Gamma^r_{rr}+2f\Gamma^r_{tr}+f^2\Gamma^r_{tt}\right) \dot r^2 = 0.
    \label{equazione_radiale_delle_geodetiche_senza_trasformazione}
\end{equation}
Expressing the Christoffel symbols in terms of the metric $g_{\mu\nu}$, 
we finally get the explicit form of the differential equation 
(\ref{equazione_radiale_delle_geodetiche_senza_trasformazione}), 
\begin{equation}
    \ddot r = {\frac{r{H_t}\dot r^2}{\sqrt{1-{\frac{2M}{r}}}\left(\sqrt{1-{\frac{2M}{r}}}-Hr\right)^2}},
    \label{espressione_finale_compatta_articolo_equazione_radiale}
\end{equation}
which is in agreement with \cite{MVLegacy}. 
Now in order to solve equation (\ref{espressione_finale_compatta_articolo_equazione_radiale}), we consider a massless particle closed to $r = 2M$ and we introduce the new coordinate $R$ defined by $r = R + 2M$, with $R\rightarrow0$ as $r \rightarrow 2M$. 
%
%
Making the replacement $r = R + 2M$, we get:
\be
\ddot R = \frac{\left(R + 2M\right)H_t}{\sqrt{1-\frac{2M}{R+2M}}\left\{\sqrt{1-\frac{2M}{R+2M}}-H\left(R+2M\right)\right\}^2}\dot R^2.
\label{tuttaR}
\ee
Hence, if we notice that 
\be
\frac{2M}{2M+R} = \frac{2M}{2M\left(1+\frac{R}{2M}\right)} = \frac{1}{1+\frac{R}{2M}}
\quad 
\xrightarrow{R  \rightarrow  0} \quad 1-\frac{R}{2M} \quad \Longrightarrow \quad 
\sqrt{1-\frac{2M}{2M+R}}\sim \sqrt{1-1+\frac{R}{2M}} = \sqrt{\frac{R}{2M}} \, , 
\ee
we can simplify the right hand side of (\ref{tuttaR}) to:
\be
\ddot R = \frac{\left(R+2M\right)H_t}{\sqrt{\frac{R}{2M}}\left\{\sqrt{1-\frac{2M}{R+2M}}-H\left(R+2M\right)\right\}^2}\dot R^2 \quad \xrightarrow{R\rightarrow 0} \quad 
\frac{2MH_t}{\sqrt{\frac{R}{2M}}\left(-2MH\right)^2}\dot R^2 
\, .
\ee
Finally, for $R\rightarrow0$, (\ref{tuttaR}) reads:
\be
\ddot R = \frac{H_t}{H^2}\frac{\dot R^2}{\sqrt{2M}\sqrt{R}} \, .
\label{near2M}
\ee
The ratio $H_t/H^2$ in front of the right hand side of the above equation is constant because for $a(t)=t^\alpha$ and $H = a'/a$, we obtain: 
\be
\frac{H_t}{H^2} = {\frac{\alpha(\alpha - 1)}{\alpha^2}} - 1  = - \frac{1}{\alpha}. 
\ee
Therefore, there is no dependence on $t$ in (\ref{near2M}), which we can rewrite as: 
\begin{equation}
    \ddot R =  {C}{\frac{\dot R^2}{\sqrt{R}}} \, , \qquad \mbox{where} \quad C = -\frac{1}{\alpha \sqrt{2M}}, 
    \label{equazione_no_rescaling_approssimata_meglio}
\end{equation}
the above equation can be solved as follows. We define:
\be
\dot R^2 = w , 
\ee 
and compute the derivative with respect to $\lambda$ of both sides of the above equation,
\be
&& {\rm LHS} \, :  \quad \frac{d}{d\lambda}\left( \frac{dR}{d\lambda}\right)^2 = 2\dot R \ddot R \, , \nonumber \\
&& {\rm RHS} \, : \quad 
\frac{dw}{d\lambda} = \frac{dw}{dR} \frac{dR}{d\lambda} \equiv w' \frac{dR}{d\lambda} = w' \dot{R} \, , 
\label{stepsDE}
\ee
where here prime $'$ stays for the derivative with respect to $R$ to simplify the notation.
By comparing the two results we get:
\be
w' = 2  \ddot R \quad \Longrightarrow \quad \ddot R = \frac{w'}{2} \, , 
\label{wprime}
\ee
which we replace in \eqref{equazione_no_rescaling_approssimata_meglio} to finally get a first order differential equation for $w(R)$, namely 
\be
\frac{w'}{2} = \frac{C}{\sqrt{R}}w \, ,
\ee
whose solution integrating in $R$ is:
\be
w(R) = B \, e^{4 C \sqrt{R} } \, , 
\ee
where $B$ is the integration constant. Since $w = \dot R^2$, 
the original equation simplifies to:
%
%
\be
\left( \frac{dR}{d\lambda}\right)^2 = B \, e^{4C\sqrt{R}} \quad \Longrightarrow \quad \frac{dR}{d\lambda} = \pm \sqrt{B}e^{2C\sqrt{R}},
\label{LastR}
\ee
Since we are dealing with ingoing geodesics, we choose the minus sign. Integrating (\ref{LastR}) for small $R$,
\begin{equation}
    R(\lambda) = -k\lambda + \rm const. \, ,
    \label{soluzione_R_lambda_no_rescaling}
\end{equation}
where $k = \sqrt{B}$. Hence, in order to reach $R = 0$, or equivalently $r = 2M$, a massless particle needs a finite amount of the affine parameter $\lambda$. Therefore, the metric is {\em geodetically incomplete}.

 
 A comment on quantum gravity, too often mentioned without a sound reason, is in order. 
 Many articles and books often justify the need for a theory of quantum gravity because of singularities at very small scales, namely beyond the Planck scale, where, it is said, Einstein's gravity breaks down. In this section, we have shown the groundlessness of such statement. Indeed, the energy scale at which the singularity occurs for the McVittie spacetime is macroscopic and not microscopic, it is in $2M$ and not in zero. 
 Therefore, singularities are not the sign of the inevitability of considering quantum effects, but actually they tell us about the presence of the hidden Weyl symmetry. 
 Even worse, all classical theories are trivially conformally invariant and thus singularity-free. The problem of singularities pop up if the quantum theory is anomalous, namely the effective quantum action explicitly breaks conformal invariance. Therefore, quantum mechanics does not solve the problem of singularities, but is instead the reason of them if the theory is not finite.

\section{Collapsing matter in an expanding Universe}

In the previous section we considered the case of a constant mass $M$ in an expanding Universe. Hence, one could question whether the singularity could actually be a pathology of the static solution for the compact object. In order to shed light on this issue, we propose a generalization of the Vaidya gravitational collapse to the McVittie-Vaidya metric.
%
Let us briefly recap the case of the Schwarzschild metric \cite{Fabbri}. We introduce the radial tortoise coordinate and make the following coordinates transformation, 
\be
r^* = r + 2M\, \ln\frac{|r - 2M|}{2M} \, , \quad t = v - r^* = v - 2M\cdot \ln\frac{|r - 2M|}{2M} - r \, . 
\ee

Hence, differentiating the time-like coordinate $t$:
\be
dt = dv + dr\left( \frac{r}{2M - r}\right) \, ,
\label{dt_for_Vaidya}
\ee
and replacing the latter into the Schwarzschild metric: 
\begin{equation}
    ds^2 = -\left(1-\frac{2M}{r}\right)dt^2 + \frac{dr^2}{1-\frac{2M}{r}} + r^2d\Omega^2 \, , 
    \label{Schwarzschild}
\end{equation}
%
%
we get (\ref{Schwarzschild}) in the coordinates $(v,r,\theta, \phi)$, 
\begin{equation}
    ds^2 = -\left(1-\frac{2M}{r}\right)dv^2 + 2drdv + r^2d\Omega^2 \, .
    \label{SchwarzschildVaidya}
\end{equation}
We obtain the Vaidya metric by replacing $M$ with the dynamical mass $M(v)$ in (\ref{SchwarzschildVaidya}). 

Now we implement the same change of coordinates to the McVittie metric. Specifically, a substitution of \eqref{dt_for_Vaidya} in \eqref{metrica_di_McVittie_con_il_raggio_areale} yelds:
\be
ds^2 = &&-\left(1-\frac{2M}{r}-H^2r^2\right)\left\{dv^2 + \frac{dr^2}{\left(1-\frac{2M}{r}\right)^2}-2\frac{drdv}{1-\frac{2M}{r}}\right\} +\\ 
&&-\frac{2Hr}{\sqrt{1-\frac{2M}{r}}}dr\left(dv - \frac{dr}{1-\frac{2M}{r}}\right) + \frac{dr^2}{1-\frac{2M}{r}} + r^2d\Omega^2.
\ee
After some algebra, and promoting $M$ to $M(v)$, one gets:
\begin{equation}
\begin{split}
     ds^2 =& -\left(1-\frac{2M(v)}{r}-H^2r^2\right)dv^2 + 2\left( 1-\frac{H^2r^2}{1-\frac{2M(v)}{r}}-\frac{Hr}{\sqrt{1-\frac{2M(v)}{r}}}\right) drdv+ \\&+ \left[ \frac{H^2r^2}{\left(1-\frac{2M(v)}{r}\right)^2}+ \frac{2Hr}{\left(1-\frac{2M(v)}{r}\right)^{\frac{3}{2}}}\right] dr^2 + r^2d\Omega^2.
    \label{McVittie-Vaidya}
    \end{split}
    \end{equation}
We can check the correctness of the result by putting $H = 0$: in this case, the metric becomes equal to \eqref{SchwarzschildVaidya}.

A computation of the Ricci scalar $R$ shows a singularity in $2M(v)$, which is now time dependent but still exactly in the same location:
{
\be
R \sim \frac{1}{ [r-2M(v)]^4 } \, .
\ee
}
Therefore, the singularity issue in $2M$ seems independent on the details of the collapse. 



\section{Avoiding the spacetime singularities in conformal gravity}
In the last fifteen years, we have finally understood that the real prediction of quantum gravity, regardless of the specific details of the theory, is the realization that the laws of physics respect the Weyl conformal invariance
at classical as well as at quantum level 
\cite{Krasnikov, kuzmin, modesto}. 
The latter statement follows from the ultraviolet finiteness of quantum gravity 
\cite{modestoLeslaw, review, Calcagni:2023goc, Calcagni:2024xku} consistently with unitarity \cite{Briscese:2018oyx, Briscese:2021mob}, causality \cite{scattering, causality, Briscese:2019twl, Zhao:2023tox} and perturbative stability \cite{Calcagni:2017sov, Calcagni:2018pro, StabilityMinkAO, StabilityRicciAO}. Properly defined theories have the above properties even in the presence of matter 
\cite{Universally, FiniteGaugeTheory, Modesto:2021okr, Modesto:2024qzl}. 
Similar properties are shared by local higher derivatives theories \cite{shapiro3, Modesto:2015ozb, Modesto:2016ofr, Rachwal:2021bgb, Liu:2022gun} with a correct quantum prescription \cite{Anselmi:2025uzj, Anselmi:2017yux, Anselmi:2017lia, Anselmi:2018kgz, Anselmi:2021hab, Anselmi:2022qor, Anselmi:2017ygm}. 
Therefore, the singularity issue \cite{SC,Modesto:2019cvh, Zhou:2019hqk, Zhang:2018qdk, Bambi:2017ott, Bambi:2017yoz, Modesto:2017uji, Bambi:2016yne}, the problems of the cosmological standard model
\cite{Calcagni:2022tuz, Modesto:2022asj}, 
and even the mystery of dark matter \cite{Modesto:2021yyf} are naturally solved by the Weyl invariance. 
Furthermore, the ambiguity in the choice of the conformal vacuum allows for stable spacetime shortcuts in our Universe \cite{Cadoni:2025cmf}. 
Aware that all the properties we will discuss in this section are common to all conformally invariant theories having the metric of interest 
as an exact solution, we will limit ourselves to consider Einstein's conformal 
invariant theory.
Indeed, Einstein's theory of gravity (EG) is actually accidentally Einstein's conformal gravity (ECG) in its spontaneously broken phase of the Weyl's symmetry. Let us briefly expand on
the above very well known statement. 
The fundamental action describing ECG in a D-dimensional spacetime is obtained through the replacement of the metric $g_{\mu\nu}$ with $\hat g_{\mu\nu}$, defined as 
\be
g_{\mu\nu} = \phi^2 \, \hat g_{\mu\nu},
\label{replace}
\ee
in the Einstein-Hilbert action 
\begin{equation}
S_{\rm EH} = \frac{2}{k_D^2}\int d^Dx\sqrt{-g}R(g),
\label{EHA}
\end{equation}
where $2/k_D^2 = 1/16\pi$ (for $c = G = 1$) and $\phi$ is the \textit{dilaton}.

After the replacement of (\ref{replace}) in (\ref{EHA}), it is guaranteed the invariance of the action under the following Weyl conformal transformation,
\be
\hat g'_{\mu\nu} = \Omega^2(x)\hat g_{\mu\nu}~,~~~~ \space \phi' =\Omega^{-1}(x) \phi \, .
\ee
 
Therefore,  any rescaling of the metric $\hat g_{\mu\nu}$, with a non trivial profile for the dilaton $\phi$, is also an exact solution of Einstein's equations, viz.
\begin{equation}
    \hat g_{\mu\nu}^* = S^2(x)\hat g_{\mu\nu}~,~~~~ \space\space \phi^* = S^{-1}(x)\phi,
    \label{tensore_metrico_riscalato_simboli}
\end{equation}
with $S(x)$ as a specific rescaling representing a selected vacuum in the broken phase of conformal symmetry.


As in other examples in the literature 
 (see for instance the already mentioned \cite{SC}), we can wisely make use of the rescaling to turn a spacetime with curvature singularities into a singularity-free and geodetically complete one, i.e., one in which an infinite amount of proper time (or of the affine parameter if we deal with massless particles) is needed to reach those singularities. 

In the following, we will limit to the case $D = 4$ and we will address the singularity in $r = 2M$. The outcome will be a solution in which $2M$ is pushed at the boundary of the Universe we live in by the means of a conformal transformation. The solution is not unique because there are multiple choices for the conformal rescaling that implement the same qualitative change to the metric. A similar rescaling was introduced for the first time in \cite{conformalons} (see also \cite{Chakrabarty:2017ysw}) and the spacetime was called \textit{conformalon}.

\subsection{Unattainability of the singularity in $2M$} 
The goal of this section is to perform a conformal rescaling of the McVittie metric with the aim to make $2M$ an unattainable point in spacetime. In other words, we will show that in the new metric a photon can never reach $2M$.
%

We repeat exactly the same analysis done in section (\ref{geoin}), 
but this time with the metric \eqref{metrica_di_McVittie_con_il_raggio_areale} rescaled by a generic conformal pre-factor $S(r)$. 
In the new metric, the radial geodesic equation in the limit $R\rightarrow0$ reads:
\begin{equation}
    \ddot R + \left( \frac{C}{\sqrt{R}}+\frac{S'(r)}{S(r)}\right) \dot R^2 = 0 .
    \label{equazione_con_il_rescaling_non_specificato}
\end{equation}
Interestingly, the contribution of the rescaling adds to the previous result in (\ref{geoin}), namely to (\ref{equazione_no_rescaling_approssimata_meglio}). 
%
In order to make the affine parameter infinite for $R \rightarrow  2M$, the term involving 
$S(r)$ in (\ref{equazione_con_il_rescaling_non_specificato}) has to scale with a power at the denominator greater than $1/2$. One simple choice could be: 
\begin{equation}
    S(r) = \left({\frac{1}{1-{2M\over r}}}\right)^n .
    \label{fattore_conforme_generico}
\end{equation}
Given the result for the first part of equation \eqref{equazione_con_il_rescaling_non_specificato} obtained in 
(\ref{geoin}), we are entitled to focus on the second term only. 
Using the conformal factor \eqref{fattore_conforme_generico}, we get: 
\be
\frac{S'(r)}{S(r)} = -n\left(1-\frac{2M}{r}\right)^{-n-1}\left(\frac{2M}{r^2}\right)\left(1-\frac{2M}{r}\right)^n = -n\left(\frac{2M}{r^2}\right)\left(1-\frac{2M}{r}\right)^{-1} = -\frac{2Mn}{r\left(r-2M\right)} \, .
\label{SpS}
\ee
Defining again $r = R + 2M$, we find:
\be
\frac{S'(R)}{S(R)} = -\frac{2Mn}{\left(R + 2M\right)R} \quad \xrightarrow{R\rightarrow 0} \quad  -\frac{n}{R} \, .
\ee
%
%
%
Thus, the differential equation to solve in the limit $R \rightarrow 0$ takes the form: 
\be
\ddot R = \frac{n}{R}\dot R^2.
\label{toS}
\ee
In order to solve (\ref{toS}), we implement the same procedure as the one which led to \eqref{soluzione_R_lambda_no_rescaling}, thus, we just illustrate the main steps.
Making the same replacement ${\dot R}^2 = w$, and following the procedure in (\ref{stepsDE}), 
we obtain an equation very similar to (\ref{wprime}), namely 
\be
\frac{w'}{2} = \frac{n}{R}w \quad \Longrightarrow \quad \frac{d w}{w} = 2 n \frac{dR}{R} \, , 
\ee
which we can integrate to:
 \be
 \ln (w) = 2n  \, \ln (R) + {\rm const.} \equiv { \ln } (w) = 2n \, {\ln }(R) + \log (C) = 
 {\ln }(R^{2n} C ) 
  \, , 
  \label{wdue}
 \ee
 with $C$ another integration constant. Replacing $w$ with ${\dot R}^2$ in (\ref{wdue}) we get:
 %
\be
\dot R^2 = CR^{2n} \quad \Longrightarrow \quad \frac{dR}{d\lambda} = -\sqrt{C}R^n,
\ee
where the minus sign has been chosen because we are considering ingoing geodesics. If we redefine 
$\sqrt{C} = k$ and integrate, we find at last:
\begin{equation}
    k\lambda = -{\frac{1}{1-n}}{\frac{1}{R^{n-1}}}+ {\rm const}.
    \quad \Longrightarrow \quad 
\lambda    \equiv   - \frac{1}{k(1-n)} \left[ \frac{1}{R^{n-1}} - \frac{1}{R_0^{n-1}} \right] ,
    \label{integrale_generale_fattore_conforme_generico}
\end{equation}
where we chose $R_0$ such that $\lambda = 0$ for $R=R_0$. 

\begin{figure}[h]
    \centering
    \includegraphics[width=0.5\linewidth]{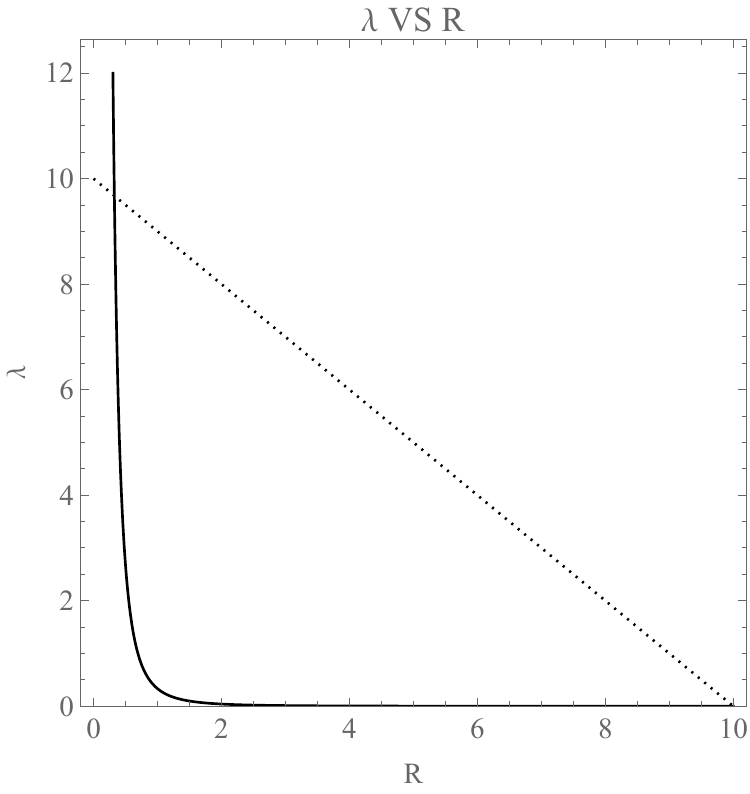}
    \caption{Plot of the affine parameter $\lambda$ as a function of $R$. The dashed line represents the solution of the geodesic equation without rescaling, whereas the continuous line represents the solution with the lowest (integer) power $n=2$ in $S(r)$ of the form \eqref{fattore_conforme_generico} ($n = 2$). Since the latter diverges as $R$ tends to 0, it comes that now the spacetime is geodesically complete in $r = 2M$. ${\rm const.} = 10$}
    \label{figure_affine_parameter}
\end{figure}

In order to get a geodetically complete spacetime the lowest (integer) power $n$ is $n = 2$. In Fig.\ref{figure_affine_parameter} is represented the affine parameter as a function of $R$, without rescaling (dashed line, eq. \eqref{soluzione_R_lambda_no_rescaling}) and with $S(r)$ of the form \eqref{fattore_conforme_generico} with $n = 2$ (continuous line, eq. \eqref{integrale_generale_fattore_conforme_generico}). 
%
%
%

One could worry that finding $S(r)$ is just half of the job, for the metric is time dependent. But when we did the first steps leading to equation \eqref{equazione_no_rescaling_approssimata_meglio}, we found this dependency to disappear. 
It deserves to be noticed that 
in equation \eqref{equazione_con_il_rescaling_non_specificato} there is no dependence on the time-like coordinate $t$ anywhere 
because the conformal rescaling only depends on the radial coordinate, and 
%
in the limit $r\rightarrow2M$ all the quantities that show a $t$-dependency cancel out. 

Since the McVittie metric shows two singularities, a rigorous treatment of the commutator between the two limits $t\rightarrow0$ and $r\rightarrow2M$ is needed. 
We define by $I$, which stays for \textit{Intra}, the quantity in brackets in \eqref{equazione_radiale_delle_geodetiche_senza_trasformazione}. Its exact expression, for $H = \alpha/t$ and $\dot H = -\alpha t^{-2}$, is:
\begin{equation}
\begin{split}
    I = \frac{\sqrt{1-\frac{2M}{r}}r^2\left(-\alpha t^{-2}\right)}{(2M-r)\left(\sqrt{1-\frac{2M}{r}}-\alpha r t^{-1}\right)^2} +\frac{\left(2M - r + 2r^2\alpha t^{-1}\sqrt{1-\frac{2M}{r}}-r^3\alpha^2 t^{-2}\right)}{r\left(\sqrt{1-\frac{2M}{r}}-\alpha r t^{-1}\right)^2}\frac{S'(r)}{S(r)}.
    \label{F}
\end{split}
\end{equation}
Performing the limit for $R\rightarrow0$ first and the limit $t\rightarrow0$ afterwards, 
\begin{equation}
    \lim_{t\rightarrow0}  \left( \lim_{R\rightarrow0}I \right)= \frac{C}{\sqrt{R}}+\frac{S'}{S} \, ,
     \label{limite_r_2M_prima_del_limite_t_0}
\end{equation}
because there is no dependence upon $t$ after the limit $r\rightarrow0$ is performed. Making now the limit $t\rightarrow 0$ first and $R\rightarrow 0$ afterwards, 
\begin{equation}
    \lim_{t\rightarrow0}I = \frac{1}{\alpha r}\frac{1}{\sqrt{1-\frac{2M}{r}}}+\frac{S'}{S},
    \label{limite_t_0_prima_del_limite_r_2M}
\end{equation}
where the infinitesimals of lower order have been neglected. 
%
 Making the replacement $r = R + 2M$ into \eqref{limite_t_0_prima_del_limite_r_2M} along with the limit $R \rightarrow 0$ yields:
\begin{equation}
   \lim_{R \rightarrow0}  \left( \lim_{t\rightarrow 0} I  \right) = \frac{1}{\alpha}\frac{1}{\sqrt{2M}\sqrt{R}}+\frac{S'}{S}.
    \label{F_quando_R_0_dopo_aver_fatto_t_0}
\end{equation}
We remind that $C = - 1/(\alpha\sqrt{2M})$, thus, for the original McVittie metric, the two limits do not commute due to a minus sign in the first term of \eqref{F_quando_R_0_dopo_aver_fatto_t_0}.
However, the discontinuity is not present for the rescaled metric because the term ${S'}/{S}$ dominates over the others in the limit $R \rightarrow 0$, as we explicitly asked while performing the rescaling.

We see that there is no dependency upon the coordinate $t$, and, therefore, there is no need to seek a conformal factor $S(r, t)$ that depends on $t$ too.

\subsection{Removing the Big-Bang singularity} 
Until now we mainly focused on the singularity in $r = 2M$. In this section we address the singularity issue in $t = 0$. We select a conformal rescaling of the following form,
\begin{equation}
S(r,t) = 1+\frac{r^2}{\left(r-2M\right)^2}\frac{\sigma^2}{t^2},
\label{Srt}
\end{equation}
where $\sigma$ is a constant having dimensions of length, so that the ratio $\sigma^2/t^2$ is dimensionless. The function (\ref{Srt}) of the variables $r$ and $t$ meets the following conditions, 
\begin{equation}
\begin{cases}
  S^{-1}(t\rightarrow\infty)& \rightarrow      \quad 1 \quad    \xrightarrow{r\rightarrow\infty}  \quad  1 \, , \\
       S^{-1}(t\rightarrow0)&  \rightarrow     \quad  0 \quad   \xrightarrow{r\rightarrow 2M}     \quad  0 \, , \\
  S^{-1}(r\rightarrow\infty)& \rightarrow     \quad  \frac{1}{1+\frac{\sigma^2}{t^2}} 
  \quad \xrightarrow{t\rightarrow\infty}\quad 1 \,  , \\
  S^{-1}(r\rightarrow  2M)& \rightarrow \quad 0 \quad  \xrightarrow{t\rightarrow0} \quad 0 \, . 
\end{cases}
\end{equation}
After the metric line element \eqref{metrica_di_McVittie_con_il_raggio_areale} is rescaled by (\ref{Srt}), the curvature invariants are regular in $2M$ and $t=0$. 
As an example, we display the Kretschmann invariant in the limit $t\rightarrow0$, and $r\rightarrow2M$. In order to distinguish the curvature invariant for the rescaled metric from \eqref{Kre}, we make use of the label 
$^* \space$ introduced for the metric after rescaling in \eqref{tensore_metrico_riscalato_simboli}.
The limit $t \rightarrow 0$ is:
\begin{align*}
    \hat R_{\alpha\beta\gamma\delta}^*\hat R^{*\alpha\beta\gamma\delta}~~\xrightarrow{~~t\rightarrow0~~}~~ & \frac{3840 \alpha^4 M^4}{\sigma^4 r^4}-\frac{3840 \alpha^4 M^3}{\sigma^4
   r^3}+\frac{1536 \alpha^4 M^2}{\sigma^4 r^2}-\frac{288 \alpha^4
   M}{\sigma^4 r}+\frac{24 \alpha^4}{\sigma^4}+\frac{4416 \alpha^3 M^3
   \sqrt{1-\frac{2 M}{r}}}{\sigma^4 r^3}+\\&-\frac{3360 \alpha^3
   M^2 \sqrt{1-\frac{2 M}{r}}}{\sigma^4 r^2}-\frac{96 \alpha^3
   \sqrt{1-\frac{2 M}{r}}}{\sigma^4}+\frac{936 \alpha^3 M
   \sqrt{1-\frac{2 M}{r}}}{\sigma^4 r}-\frac{3864 \alpha^2
   M^3}{\sigma^4 r^3}+\\&+\frac{3948 \alpha^2 M^2}{\sigma^4
   r^2}-\frac{1296 \alpha^2 M}{\sigma^4 r}+\frac{144
   \alpha^2}{\sigma^4}-\frac{720 \alpha M^2 \sqrt{1-\frac{2
   M}{r}}}{\sigma^4 r^2}-\frac{96 \alpha \sqrt{1-\frac{2
   M}{r}}}{\sigma^4}+\\&+\frac{552 \alpha M \sqrt{1-\frac{2
   M}{r}}}{\sigma^4 r}+\frac{96 M^2}{\sigma^4 r^2}-\frac{96
   M}{\sigma^4 r}+\frac{24}{\sigma^4}.
\end{align*}
The limit for $r\rightarrow2M$ is a positive constant:
${24\alpha^4}/{\sigma^4}$.
We get the same result if we compute the limit $r\rightarrow2M$ before the limit $t\rightarrow0$. 
In other words, 
the two limits, $r\rightarrow2M$ and $t\rightarrow0$, commute.

\section{Engineering the McVittie spacetime to remove the singularities}
We propose here an \textit{ad hoc} metric that avoids the singularity in $2M$ replacing $M$ with a function of the radial coordinate $M(r)$ \cite{Modesto:2010uh, Bambi:2016wmo, Burzilla:2020bkx, Burzilla:2020utr, Zhou:2023lwc, Burzilla:2023xdd, dePaulaNetto:2023cjw, Zhou:2022yio, Mo:2022szw, dePaulaNetto:2021axj,Giacchini:2021pmr}. The drawback is that the energy conditions will be all violated. 
One simple choice for the function $M(r)$ could be:
\begin{equation}
    M(r) = M\frac{r^2}{r^2+\left(bM\right)^2} .
    \label{M(r)}
\end{equation}
In order to remove the singularity in $2M$ we have to impose $ - g_{00}(r)$ to be always positive. Therefore, 
the dimensionless parameter $b$ needs to be worth at least $2$, namely $b \gtrsim 2$. 

Once again we stress that the modification needed to remove the singularity is order $M$ and not order Planck mass. This is a further confirmation that the singularity issue is not a quantum gravity problem as usually stated in literature. 

We point out that if there exist a purely gravitational theory of which (\ref{M(r)}) is a solution, then the fundamental scale of such a theory will be very large, that is, of the order of the mass $M$. Furthermore, for each different black hole mimicker we will need a different theory. These two issues seriously hamper the viability of this theoretical proposal. Therefore, a different and likely non-perturbative mechanism is needed to justify the profile (\ref{M(r)}). 
Conversely, if the theory enjoys conformal symmetry, then the scale at which the metric is modified through global rescaling can be completely arbitrary.

Finally, we compute the Misner-Sharp (MS) mass in order to see whether there is cosmological coupling between the compact object and the cosmological embedding \cite{Cadoni:2023lqe}, 
\begin{equation}
M_{\rm MS} = \frac{r}{2}\left( 1-g^{\mu\nu}\nabla_\mu r\nabla_\nu \right) = \frac{r}{2}\left( 1-g^{rr}\right)  \,\, , \,\,g^{rr} =1-\frac{2M(r)}{r}-H^2r^2\,.  
    \label{MS_mass}
\end{equation}

It follows that
\begin{equation}
M_{\rm MS} = \frac{Mr^2}{r+\left(2M\right)^2} + \frac{H^2r^3}{2}\, .
    \label{expression_MS_mass}
\end{equation}
The result of above (\ref{expression_MS_mass}) shows that there is no coupling between $M$ and $a(t)$ regardless  
of the singularity resolution. In general, we can have McVittie-like regular spacetimes violating the energy conditions that may or may not show a cosmological coupling \cite{Cadoni:2023lqe}. 


\subsection{Violation of the energy conditions}
\begin{figure}[h]
    \centering
    \includegraphics[width=0.4\linewidth]{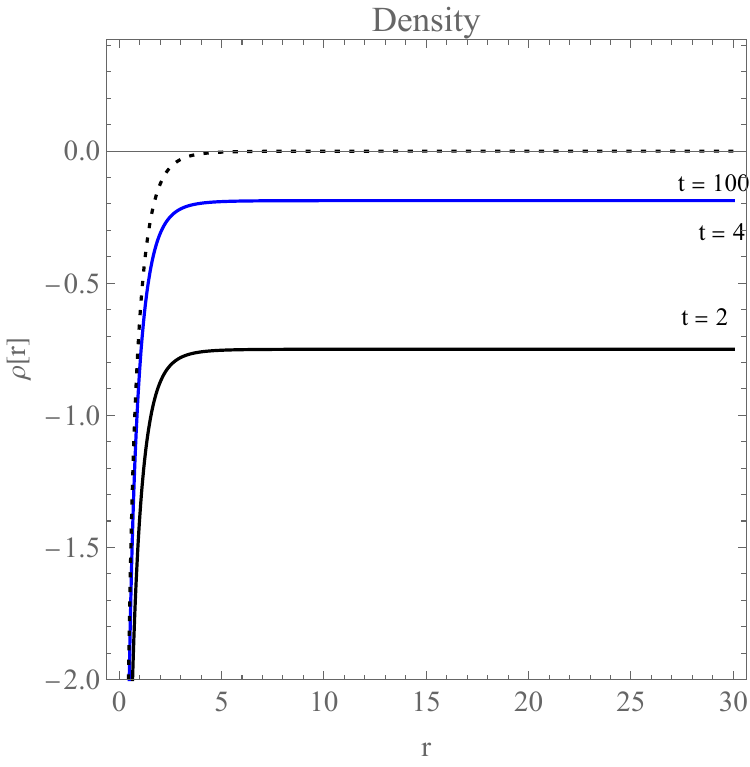}
    \includegraphics[width=0.4\linewidth]{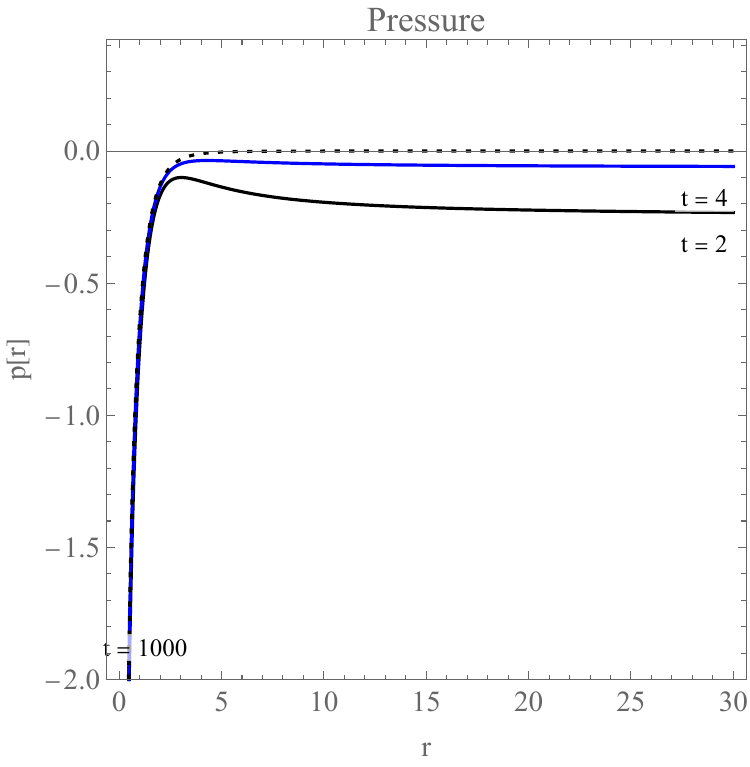}
    \includegraphics[width=0.4\linewidth]{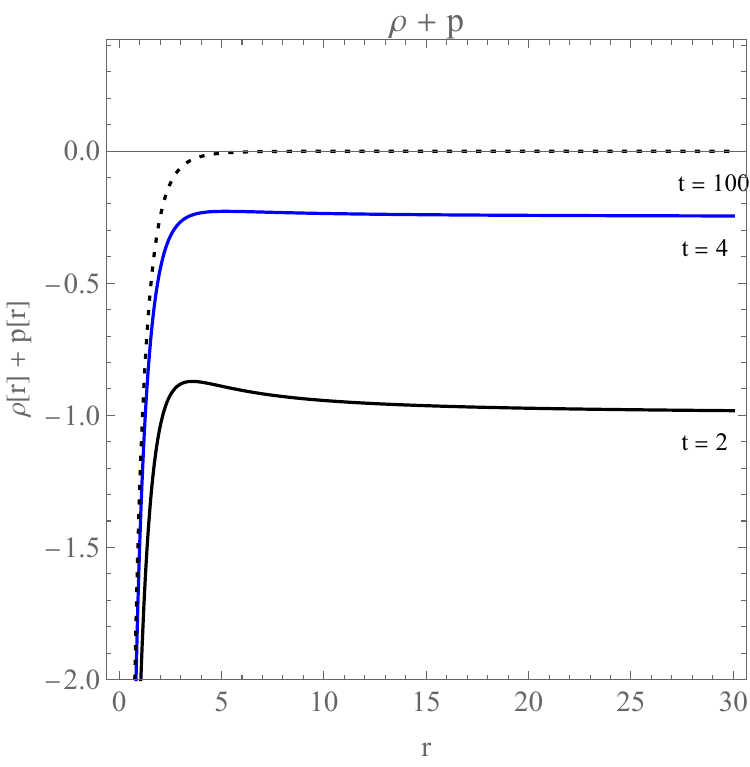}
    \includegraphics[width=0.4\linewidth]{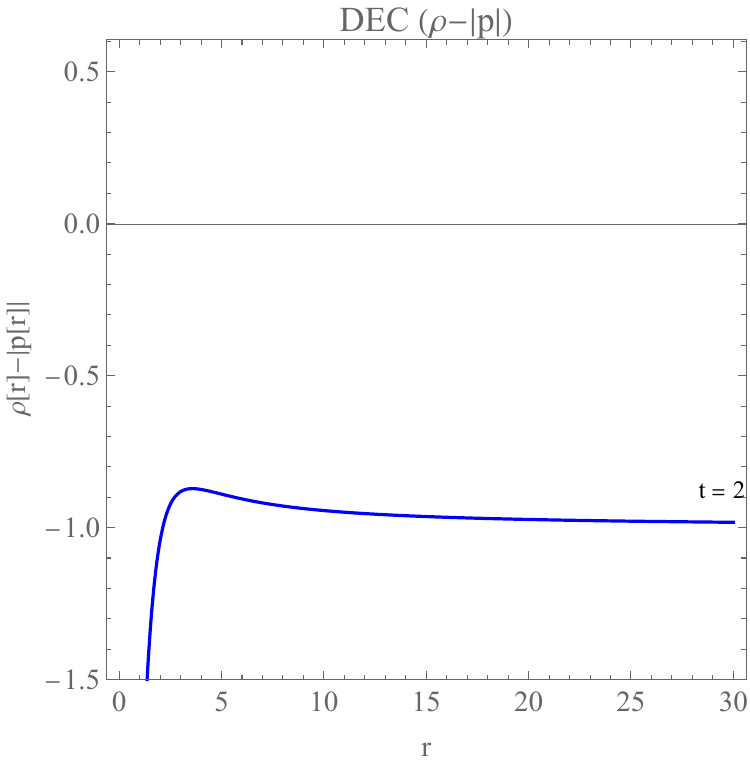}
    \caption{The first two plots at the top show the energy density and the pressure as functions of the radial coordinate $r$, for three different values of $t$, for a $M(r)$ shaped like \eqref{M(r)} with $M=1$ and $b=2$. The two plots at the bottom show the NEC (or WEC, if we combine the condition $\rho + p\ge0$ with the requirement that $\rho\ge0$) and the DEC are not satisfied. 
    }
    \label{figure_energy_conditions_with_M(r)_simpler}
\end{figure}

A look at the plots for $p$, $\rho$, and some other combinations confirms that the replacement of $M(r)$ into \eqref{metrica_di_McVittie_con_il_raggio_areale} gives rise to the violation of all the energy conditions. 
This is explicitly shown in Fig.\ref{figure_energy_conditions_with_M(r)_simpler}. We here do not display the analytic quantities because very cumbersome. 

\subsection{Apparent horizons, conformalons, and giant solitons} 

In this section, we investigate the presence of null surfaces or apparent horizons and link the analysis to the results we showed in the previous sections, namely the presence of a curvature singularity in $r = 2M$, and the violation of the DEC beneath $r = 8 M/3$.
The equation identifing the null surfaces reads: 
\begin{equation}
    g^{\mu\nu} \left(\partial_{\mu}A\right)\left(\partial_{\nu}A\right) = 0 \, ,
    \label{apphor}
\end{equation}
where $A$ is the area of the two-sphere. 
Roaming then the path of V. Faraoni \cite{Faraoni}, we apply the definition of apparent horizon to the diagonalized form of the metric \eqref{metrica_di_McVittie_con_il_raggio_areale}. Here we review the main steps of the derivation.
We change coordinates according to $dt = F dT - \beta dr$ ($F\left(t, R\right)$ is an integrating factor), where 
\be
\beta = \frac{Hr}{\sqrt{1-\frac{2M}{r}}\left(1-\frac{2M}{r}-H^2r^2\right)}\, .
\ee

Hence the off diagonal term disappearance, and the metric in the new coordinates reads:
\begin{equation}
    \begin{split}
    ds^2 = -\left(1-\frac{2M}{r}-H^2r^2\right)F^2dT^2+\frac{dr^2}{\left(1-\frac{2M}{r}-H^2r^2\right)}+r^2d\Omega^2.
        \label{metrica_di_McVittie_diagonalizzata}
    \end{split}
\end{equation}

\begin{figure}[h]
    \centering
    \includegraphics[width=0.5\linewidth]{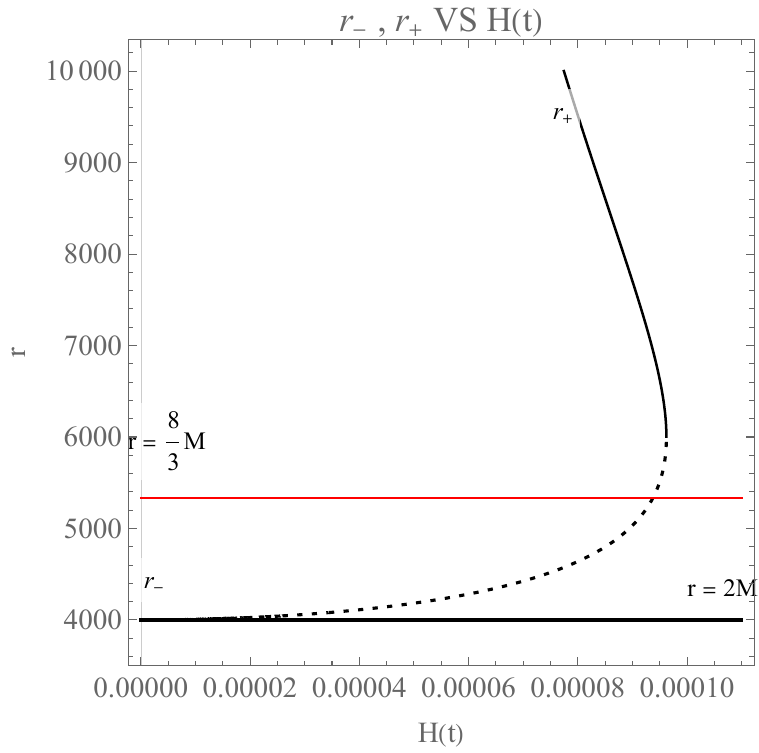}
    \caption{The two apparent horizons $r_{-}$ and $r_{+}$, along with the curvature singularity in $r = 2M$ and the radius $r = 8M/3$, are plotted as functions of the Hubble parameter $H(t)$ (here $M=2000$). Hence, the time coordinate $t$ increases from right to left. 
A thin gray vertical line near zero marks the value of $H(t)$ at the present time. 
We can infer two features: $2M$ is almost a naked singularity, and the instability region marked by 
$r = 8M/3$ is outside the horizon.} 
\label{figure_apparent_horizons}
\end{figure}

Now equation \eqref{apphor} is satisfied if $g^{rr} = 0$, that is, according to \eqref{metrica_di_McVittie_diagonalizzata},
\begin{equation}
    H^2r^3-r+2M = 0.
    \label{espressione_per_collocare_gli_orizzonti_esplicita}
\end{equation}
The above cubic equation admits three roots one of which is negative and can be discarded (negative distances are of no physical interest). 
The two other solutions are instead apparent horizons. Following the notation introduced in \cite{MVLegacy}, we indicate these solutions with $r_+$ and $r_{-}$, which define respectively the bigger and the smaller positive root. 
Specifically, when the coordinate $t$ flows, $r_{-}$ decreases, if $H$ is proportional to $\frac{1}{t}$. 
From Fig.\ref{figure_apparent_horizons}, we see that $r_-$ lowers down towards $2M$, while 
$r_+$ tends to infinity as $H\rightarrow0$.
We notice that $2M$ is not a naked singularity although according to the current value of $H$ (the thin grey vertical line plotted near zero in Fig.\ref{figure_apparent_horizons}) $2M$ is {\em almost naked}.


To make more quantitative how narrow is the region between the horizon and $2M$, we can compute the proper physical distance between the two surfaces. Since such a radial distance is measured at fixed $t$, $dt = 0$, thus
we get:
\be
\ell_r = \int \frac{dr}{\sqrt{1-\frac{2M}{r}}} = r \sqrt{1-\frac{2M}{r}}+2M\, \tanh ^{-1}\left(\sqrt{1-\frac{2M}{r}}\right).
\ee
The root $r_{-}$, that is, the apparent horizon we are interested in, reads:

\begin{equation} 
r_{-} = \frac{(-3)^{2/3} H^2-\sqrt[3]{-3} \left(\sqrt{81 H^8 M^2-3 H^6}-9 H^4 M\right)^{2/3}}{3 H^2 \sqrt[3]{\sqrt{81 H^8
   M^2-3 H^6}-9 H^4 M}}.
    \label{root_r_minus}
\end{equation}
If we expand the above expression for small $H$, we find:
\be
r_{-} \approx  2M + 8 M^3H^2.   
\ee
The distance $\Delta$ is obtained computing the following difference,  
\be
\Delta &= & r \sqrt{1-\frac{2M}{r}}+2 M \tanh ^{-1}\left(\sqrt{1-\frac{2M}{r}}\right)\Bigg|_{r = 2M + 8M^3H^2} 
– \left[ r \sqrt{1-\frac{2M}{r}}+2M \tanh ^{-1}\left(\sqrt{1-\frac{2M}{r}}\right)\right] \Bigg|_{r = 2M} 
\nonumber  \\ 
& = &\sqrt{1-\frac{2M}{8 H^2M^3+2M}} \left(8 H^2M^3+2M\right)+2M \tanh ^{-1}\left(\sqrt{1-\frac{2M}{8 H^2M^3+2M}}\right) \approx 8 H M^2 \, ,
\ee
where again the last result comes after a series expansion about H. Actually, $H^2 M^3 \ll M$. 
For a solar mass black hole $2M \approx 3 {\rm km}$, hence, since 
nowadays $H\sim 10^{-18}\rm m^{-1}$, we get 
\be
\Delta \approx 10^{-11} {\rm m} \, , 
\label{Delta}
\ee
which is of the order of magnitude of the Bohr's radius. Of course, if we increase the Schwarzschild radius the distance (\ref{Delta}) increases too.

Moreover, $r_{-}$ is beneath $8M/3$, highlighting a region of tachyonic instability outside the horizon.


The two solutions $r_{-}$ and $r_{+}$ exist only if a certain relation between the mass and the Hubble parameter is satisfied \cite{Mariano}.
%



%
Looking at the function $f(r) = H^2r^3-r+2M$, if drawn, 
it can show three intersections with the horizontal axis, two of which are in the positive domain: 
%
$r_{-}$ and $r_{+}$. 
Such roots exist if 
$f(r)$ has a stationary point somewhere between the two intersections. 
For the Rolle theorem, a function continuous in $[a, b]$, differentiable in $(a, b)$, and such that $f(a) = f(b)$ must have a stationary point (actually a minimum) somewhere in the interval $(a, b)$. Thus, we calculate the first derivative and we put it equal to zero:
\begin{equation}
    f'(r) = 3H^2r^2-1 = 0 \quad \Longrightarrow \quad r_{\rm m} = \frac{1}{\sqrt{3}H}\,, 
\end{equation}
where $r_{\rm m}$ is the position of the minimum. If the two roots exist, $f(r_{\rm m})$ has to be negative:
\begin{equation}
    f(r_{\rm m}) = H^2 r_{\rm m}^3-r_{\rm m} + 2M = 2M - \frac{2}{3 \sqrt{3} H} < 0 \quad \Longrightarrow 
    \quad 
     M < \frac{1}{3\sqrt{3}H} .
     \label{condMH}
\end{equation}
Therefore, if the above condition (\ref{condMH}) holds we have two horizons. 
At the same time, the condition \eqref{condMH} explains why the apparent horizon does not exist for small values of $t$, that is, for large $H$, for a fixed value of the mass $M$.

We elaborate now on the causal structure of the singularity-free upgraded McVittie spacetime.

\underline{{\em The McVittie Conformalon}. }
Since the causal structure is conformally invariant the trapped surfaces stay the same for the singularity-free metric (\ref{tensore_metrico_riscalato_simboli}). Namely the rescaled McVittie metric with $S(r)$ given for example by (\ref{fattore_conforme_generico}) has the horizons located in the same position that in the McVittie spacetime. 
In particular the likely unstable region defined by $r<8M/3$ is present also for the McVittie-Conformalon. 
However, here we can see the full power of conformal symmetry at work. Indeed we can replace $2M$ with $8M/3$ in (\ref{fattore_conforme_generico}) making the DEC violating region unreachable. A proper rescaling could be:
\be
 S(r) = \left({\frac{1}{1-{8M\over 3r}}}\right)^n .
 \label{8M3}
\ee
 The metric is now singularity free and all the energy conditions are satisfied in the all geodetically complete Universe, but there is no black hole.

\begin{figure}[h]
    \centering
    \includegraphics[width=0.5\linewidth]{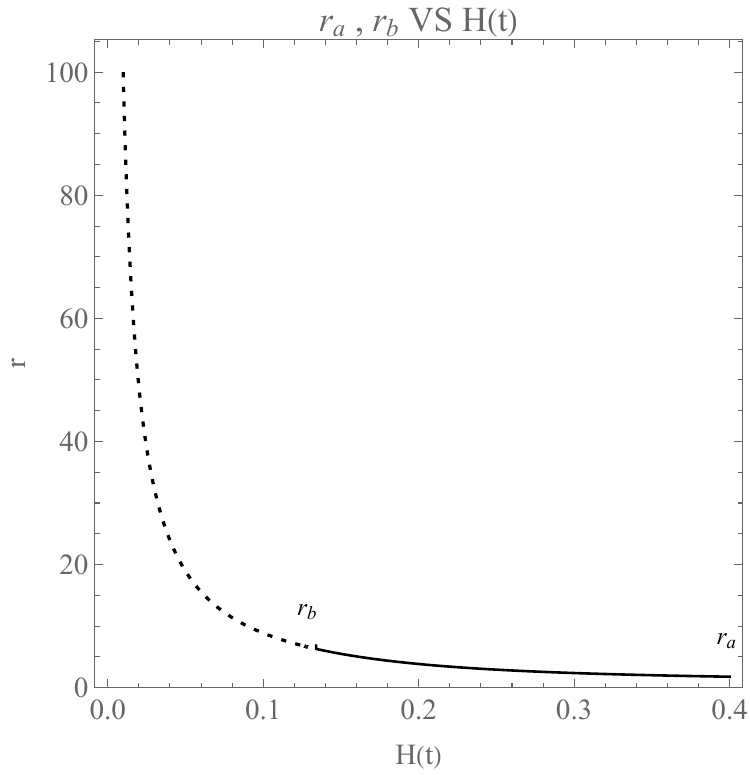}
    \caption{The cosmological horizon which results from plotting together the two solutions of \eqref{eq_app_hor_with_M(r)} that are physically relevant. The two roots are: $r_a$ and
$r_b$, for $M(r)$ given by \eqref{M(r)}. Here $M = 1$ and $b = 2$. 
}
    \label{cosmological_horizons_plot}
\end{figure}

\underline{{\em The Giant Soliton}. }
 Let us now consider what happens to the location of the apparent horizons if we replace $M(r)$ as given in \eqref{M(r)} into the line element \eqref{metrica_di_McVittie_con_il_raggio_areale}. 
 %
 %
 Now equation (\ref{apphor}) 
 becomes:
 \begin{equation}
 H^2r^3 - r  + 2\left( \frac{Mr^2  }{r^2 + \left( b M\right)^2}\right) = 0 \,\,\,\, \rightarrow \,\,\,\, H^2r^3\left[ r^2 + \left( b M\right)^2\right] - r\left[ r^2 + \left(b M\right)^2 \right] + 2Mr^2  = 0\, ,
     \label{eq_app_hor_with_M(r)}
 \end{equation}
 and it has now five roots. But only two of those are physically meaningful, for they are the only positive solutions. If we plot them together in a graph, as shown in Fig.\ref{cosmological_horizons_plot}, we see that they are two branches of a single cosmological horizon wherein we are supposed to live.
%

It turns out that removing the singularity in $2M$ turns the spacetime into a kind of soliton rather than a black hole.


\section*{Conclusions and Remarks}

We presented a detailed synthesis of the main features of the McVittie's spacetime. The metric is an exact solution of Einstein's field equations with an energy momentum tensor satisfying 
the NEC, the SEC, and the WEC, but not the DEC in the vicinity of $2M$.
%
It is crucial to notice that the McVittie metric is a very physical one whether we want an exact solution that merges a compact object in an evolving Universe. Indeed, any attempt to modify the McVittie spacetime will lead to further violate the energy conditions. Therefore, we do not see room for other solutions consistent with ordinary not exotic matter.
However, the metric points out the following serious issues: it has a curvature singularity in $2M$, which is attainable in finite affine time, and shows a violation of the DEC for $r<8M/3$. 
The latter sounds alarming because an observer would see something like a shell of a superluminal fluid outside the apparent horizon (from $8M/3$ to $2M$), a thing that casts some doubts upon the plausibility of the proposed McVittie solution, at least in such region. 
About the singularity in $2M$, it is {\em mathematically} covered by an apparent horizon, which, to date, is just an atomic radius far away from $2M$ for a solar mass black hole. The solution has another horizon, which is cosmological and lifts to $+ \infty$ as $t\rightarrow + \infty$. 
In order to model the dynamical gravitational collapse, we proposed what has been here called McVittie-Vaidya metric. It turns out that the singularity in $r = 2M$ does not move nor disappear if the mass is taken to be a function of time-like coordinate $v$ in a gravitational collapse described through the new metric. 
Hence, can we really take seriously the presence of a Schwarzschild black hole in our Universe? 
A crucial aspect to take into account whether we do not want to give a hasty and/or naive answer to this question concerns the analyticity of the McVittie solution in the derivative of the Hubble function, $H^\prime(t)$. Indeed, the singularity in $2M$ disappears only when $H(t)$ is exactly constant or $H^\prime(t) =0$, but stays there for any arbitrarily small but not zero function $H^\prime(t)$. In other words, the Schwarzschild (or Schwarzschild - (anti)de Sitter) metric is not a perturbative limit of the McVittie spacetime. 
In other words, physics on very large scales influences local physics in a non-perturbative way: the black hole and the entire Universe are entangled each other\footnote{This nonlocal feature of the McVittie Solution also seems to sound like an alarm bell also in the dark matter issue \cite{Li:2019ksm, Modesto:2021yyf}.}.
Therefore, since the Universe we live in is asymptotically FRW, or at least it appears so, we are forced to take the McVittie metric seriously whether we want to describe a compact object in an expanding Universe. 
One could think to consider the Schwarzschild metric correct in a finite region of space, a kind of void in the  Universe, but this would be unlikely to be credible because we we will not have an analytical interpolation with the FRW spacetime.
At the moment the study of a locally spherically symmetric spacetime seems to be a relatively academic exercise. Hence, in the future, we plan to extend the McVittie spherically symmetric spacetime to a rotating one that is more suitable for a connection to astrophysical observations.

In the second part of the paper, we made two proposals about the singularity resolution. 
In order to solve the singularity issue 
we proposed two scenarios: one (addressing both $r = 2M$ and $t = 0$) based on conformal symmetry that does not lead to further violations of the energy conditions, and a more exotic one (addressing $r = 2M$) eventually traceable in a nonlocal or higher derivative theory (see for example the already mentioned \cite{review}). This latter (giant soliton) consists in the choice of a mass $M(r)$ such that the singularity is no more present, with the shortcoming that the EC are not satisfied anymore. The function $M(r)$ shows a dimensionless parameter $b$ of order one consistent with a  
modification of the classical spacetime at macroscopic energy scales far below the Planck scale. In other words, EG should break down already at macroscopic scales. 
%
For this object, 
instead of the two horizons, there is only a cosmological horizon. 
An explicit computation of the Misner-Sharp (MS) mass with $M(r)$ in place of $M$ shows that for a singularity-free McVittie spacetime in general there is not a coupling between the mass and the scale factor $a(t)$. According to \cite{Cadoni:2023lqe} the cosmological coupling depends on the parametrization of the MacVittie metric we start from in order to remove the singularity.  



The former proposal (conformalons) turns the singularities in asymptotic regions of the Universe, namely nothing can reach the singularities. The latter turns the black hole in a giant soliton whose spacetime can be explored in finite time. 
Nevertheless, there is another possible scenario. We can assume the Universe to be asymptotically Minkowski and turn it in an FRW mimicker by the means of a conformal rescaling. Therefore, the singularity issue in $r=0$ can be solved like for example in \cite{conformalons, Modesto:2019cvh}, while the metric is rescaled by $a(\eta)$, where now $\eta$ is the conformal time. Explicitly:
\be
ds^2 = a(\eta) \left( 1 + \frac{L^4}{r^4} \right) 
\left[ - \left( 1 - \frac{2 M}{r} \right)d \eta^2  + \frac{dr^2}{1 - \frac{2 M}{r} } + r^2 d \Omega^{(2)} \right] \, .
\ee
The above metric is at the moment just a proposal that has to pass all the cosmological tests in order to be taken seriously from the observational point of view. However, it could become a real alternative if the observational evidences for a Schwarzschild black hole regular in $2M$ become unquestionable.

\section*{Acknowledgements}

We thank L. Orlando for helpful discussions about the energy conditions, M. Cadoni, M. Pitzalis, and A. P. Sanna for the numerous discussions and criticisms throughout the writing of the all paper, and C. Bambi for providing material useful in support of the analiticity requirement.

\end{document}